\documentclass[aps,prl,twocolumn,superscriptaddress,10pt]{revtex4-1}
\usepackage{graphicx,xcolor,longtable,physics}
\usepackage[super]{nth}
\newcommand{\Xe}{\ensuremath{^{129}\mathrm{Xe}}}
\newcommand{\He}{\ensuremath{^{3}\mathrm{He}}}
\begin{document}

\title{A new measurement of the permanent electric dipole moment of $^{129}$Xe using $^{3}$He comagnetometry and SQUID detection}

\author{N. Sachdeva}
\email{sachd@umich.edu}
\affiliation{Department of Physics, University of Michigan, Ann Arbor, Michigan 48109, USA}
\author{I. Fan}
\affiliation{Physikalisch-Technische Bundesanstalt (PTB) Berlin, 10587 Berlin, Germany}
\author{E. Babcock}
\affiliation{J\"ulich Center for Neutron Science, 85748 Garching, Germany}
\author{M. Burghoff}
\affiliation{Physikalisch-Technische Bundesanstalt (PTB) Berlin, 10587 Berlin, Germany}
\author{T. E. Chupp}
\affiliation{Department of Physics, University of Michigan, Ann Arbor, Michigan 48109, USA}
\author{S. Degenkolb}
\affiliation{Department of Physics, University of Michigan, Ann Arbor, Michigan 48109, USA}
\affiliation{Institut Laue-Langevin, 38042 Grenoble, France}
\author{P. Fierlinger}
\affiliation{Excellence Cluster Universe and Technische Universit{\"a}t M{\"u}nchen, 85748 Garching, Germany}
\author{E. Kraegeloh}
\affiliation{Excellence Cluster Universe and Technische Universit{\"a}t M{\"u}nchen, 85748 Garching, Germany}
\affiliation{Department of Physics, University of Michigan, Ann Arbor, Michigan 48109, USA}
\author{W. Kilian}
\affiliation{Physikalisch-Technische Bundesanstalt (PTB) Berlin, 10587 Berlin, Germany}
\author{S. Knappe-Gr{\"u}neberg}
\affiliation{Physikalisch-Technische Bundesanstalt (PTB) Berlin, 10587 Berlin, Germany}
\author{F. Kuchler}
\affiliation{Excellence Cluster Universe and Technische Universit{\"a}t M{\"u}nchen, 85748 Garching, Germany}
\affiliation{TRIUMF, Vancouver, British Columbia V6T 2A3, Canada}
\author{T. Liu}
\affiliation{Physikalisch-Technische Bundesanstalt (PTB) Berlin, 10587 Berlin, Germany}
\author{M. Marino}
\affiliation{Excellence Cluster Universe and Technische Universit{\"a}t M{\"u}nchen, 85748 Garching, Germany}
\author{J. Meinel}
\affiliation{Excellence Cluster Universe and Technische Universit{\"a}t M{\"u}nchen, 85748 Garching, Germany}
\author{Z. Salhi}
\affiliation{J\"ulich Center for Neutron Science, 85748 Garching, Germany}
\author{A. Schnabel}
\affiliation{Physikalisch-Technische Bundesanstalt (PTB) Berlin, 10587 Berlin, Germany}
\author{J. T. Singh}
\affiliation{National Superconducting Cyclotron Laboratory and Department of Physics \& Astronomy, Michigan State University, East Lansing, Michigan 48824, USA}
\author{S. Stuiber}
\affiliation{Excellence Cluster Universe and Technische Universit{\"a}t M{\"u}nchen, 85748 Garching, Germany}
\author{W. A. Terrano}
\affiliation{Excellence Cluster Universe and Technische Universit{\"a}t M{\"u}nchen, 85748 Garching, Germany}
\author{L. Trahms}
\affiliation{Physikalisch-Technische Bundesanstalt (PTB) Berlin, 10587 Berlin, Germany}
\author{J. Voigt}
\affiliation{Physikalisch-Technische Bundesanstalt (PTB) Berlin, 10587 Berlin, Germany}

\date{\today}

\begin{abstract}
We describe a new technique to measure the EDM of $^{129}$Xe with $^3$He comagnetometry. Both species are polarized using spin-exchange optical pumping, transferred to a measurement cell, and transported into a magnetically shielded room, where SQUID magnetometers detect free precession in applied electric and magnetic fields. The result of a one week run combined with a detailed study of systematic effects is $d_A(\Xe) = (0.26 \pm 2.33_\mathrm{stat} \pm 0.72_\mathrm{syst})\times10^{-27}~e\,\mathrm{cm}$. This corresponds to an upper limit of $|d_A(\Xe)| < 4.81\times 10^{-27} ~e\,\mathrm{cm}~(95\%~\mathrm{CL})$, a factor of 1.4 more sensitive than the previous limit.
\end{abstract}
%\keywords{}
\maketitle %consistency, due to, in all cases, checks
Searches for permanent electric dipole moments (EDMs) are a powerful way to investigate beyond-standard-model (BSM) physics. An EDM  is a charge asymmetry along the total angular momentum axis of a particle or system and is odd under both parity reversal (P) and time reversal (T).  Assuming CPT conservation (C is charge conjugation) an EDM is a direct signal of CP violation (CPV), a topic of current interest in part because it is a condition required to generate the observed baryon asymmetry of the universe~\cite{Sakharov1967}. The Standard Model incorporates CPV through the phase in the CKM matrix and the QCD parameter $\bar{\theta}$. However, the Standard Model alone is insufficient to explain the size of the baryon asymmetry~\cite{Dine2003}, motivating the search for BSM CPV. BSM scenarios that generate the observed baryon asymmetry~\cite{Morrissey2012} generally also provide for EDMs larger than the SM estimate $|d_A(\Xe) ^\mathrm{SM}|\approx 5\times 10^{-35}~e\,$cm~\cite{Chupp2017}. Additional motivation is provided by the consideration of $\Xe$ as a comagnetometer in neutron EDM experiments~\cite{Degenkolb2012,Masuda2012}, which require $d_A(\mathrm{^{129}Xe})\lesssim 3\times 10^{-28} e\,\mathrm{cm}$ in order to measure the neutron EDM with sensitivity $1\times 10^{-27} e\,\mathrm{cm}$.

Beginning with the neutron~\cite{Smith1957}, EDMs measured in several systems have provided constraints on how BSM CPV can enter low-energy physics (see~\cite{Chupp2017} for a review). EDMs of diamagnetic systems such as $\Xe$ and $^{199}$Hg are particularly sensitive to CPV nucleon-nucleon interactions that induce a nuclear Schiff moment. Diamagnetic systems are also essential for constraining electron-spin independent low-energy CPV parameters~\cite{Chupp2015}. While the most precise atomic EDM measurement is from $^{199}$Hg~\cite{Graner2016}, there are theoretical challenges to constraining hadronic CPV parameters from $^{199}$Hg alone, and improved sensitivity to the $\Xe$ EDM would tighten these constraints~\cite{Chupp2015, Yamanaka2017}. Recent work has shown that diamagnetic-atom EDM contributions from light-axion-induced CPV are significantly stronger for $\Xe$ than for $^{199}$Hg~\cite{DzubaFlambaum2018}.

The first $\Xe$ EDM measurement by Vold \textit{et al.} monitored $\Xe$ Larmor precession frequency as a function of applied electric field~\cite{Vold1984}. Rosenberry \textit{et al.}~\cite{Rosenberry2001} used a two-species maser with a $\He$ comagnetometer providing the upper limit $|d_A(\Xe)| \le 6.6 \times 10^{-27}~e\,\mathrm{cm}$ (95\% CL), the most sensitive $\Xe$ measurement to date. A number of $^{129}$Xe EDM efforts to improve on this limit have followed, including an active maser technique~\cite{Inoue2016}, and an experiment with polarized liquid xenon~\cite{Ledbetter2012}. An approach similar to ours using free precession and SQUID magnetometry is also being pursued~\cite{Schmidt2013}. 

For a system with total angular momentum $\vec F$,  EDM $d\vec F/F$, and magnetic moment $\mu\vec F/F$, the Hamiltonian is $H = -(\mu \vec F\cdot\vec B + d\vec F\cdot\vec E)/F$. This results in an energy splitting dependent on $\vec E\cdot \hat B$ and a corresponding frequency shift $\omega_d =\pm d\,|E|/(\hbar F)$ between states with $|\Delta m_F|=1$. Changes of $\vec B$ due to drifts and extraneous magnetic fields, for example from leakage currents, lead to frequency shifts that are mitigated by comagnetometry---simultaneous measurement with a colocated species. The $^{129}$Xe-$^3$He comagnetometer system is nearly ideal because both can be simultaneously polarized by spin-exchange optical pumping (SEOP) \cite{Chupp1988}, have long spin relaxation times enabling precision frequency measurements, and $\He$, with $27\times$ lower nuclear charge $Z$, is much less sensitive to CP violation~\cite{FlambaumGinges}. 

\begin{figure}[!t]
\centering
%\vskip -0.14 truein
\includegraphics[width=0.96\columnwidth]{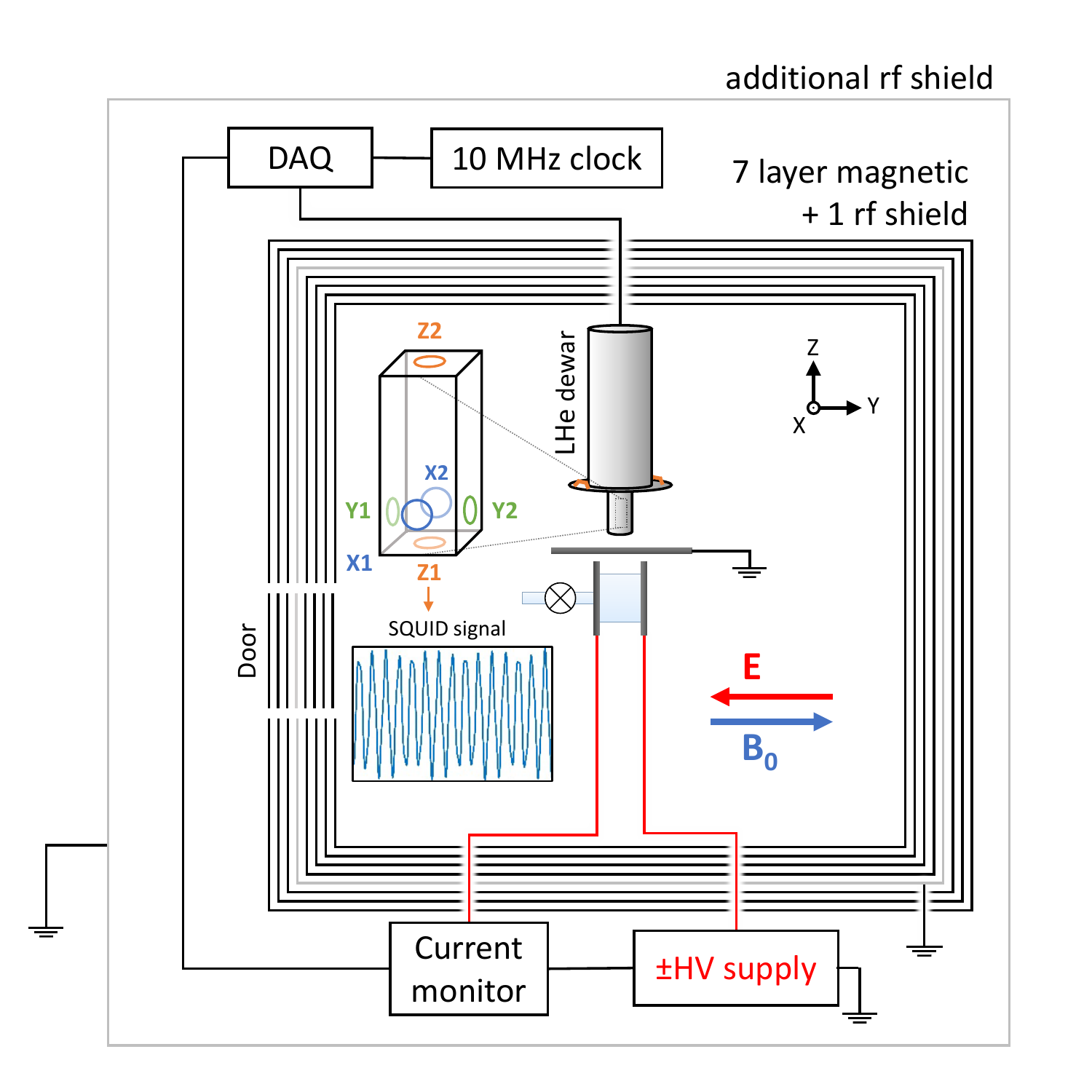}
\vskip -0.10 true in
\caption{\label{apparatus} (color online) Schematic of the HeXeEDM apparatus at PTB. The electric field $\vec E$ indicated corresponded to +HV and the magnetic field is shown along $+\hat y$. The inset shows a typical raw SQUID signal for 1/2 second of data; the frequencies are 30.8~Hz for $\Xe$ and 84.8~Hz for $\He$. Not to scale.}
\vskip -0.25 true in
\end{figure}

The layout of the HeXeEDM experiment, previously described in~\cite{Kuchler2016}, is shown in Fig. \ref{apparatus}. Free precession of the $\Xe$ and $\He$ magnetization was measured with an array of six low-noise superconducting quantum interference devices (SQUIDs) in the magnetically shielded room BMSR-2 at Physikalisch-Technische Bundesanstalt (PTB) Berlin. BMSR-2 provided a passive shielding factor of 75,000 for frequencies below 0.01 Hz and more than $10^8$ above 6 Hz~\cite{Bork2002}. A 1.6 m diameter set of Helmholtz coils generated a static magnetic field of $2.6~\mathrm{\mu T}$ along the $y$-axis. In a separate setup similar to that described in Ref.~\cite{Korchak2013}, the gas mixture of 18\% isotopically enriched xenon ($90\%$~$\Xe$), $73\%~\He$, and $9\%~\mathrm{N_2}$ was polarized by SEOP in a refillable optical pumping cell (OPC). Simultaneous polarization of $\Xe$-$\He$ mixtures compromise both polarizations because the optimum conditions are very different for the two species. Typically, we achieved 9--12\% polarization for $\Xe$ and 0.1--0.2\% polarization for $\He$ depending on the total pressure in the OPC. We used two valved EDM  cells with 30~mm diameter, 2~mm thick, p-type (Boron) doped 1-10 $\Omega$\,cm silicon electrodes diffusion bonded to borosilicate glass cylinders~\cite{PatrickPistelPC}. One cell (PP1) had a length of 18.5~mm and an inner diameter of 20.5~mm; the other cell (PP2) had a length of 21.8~mm and an inner diameter of 20.4~mm. Before each filling, the EDM cell was degaussed using a commercial bulk degausser~\cite{degausser}. The polarized gas was expanded from the OPC into the evacuated EDM cell. Each time the OPC was refilled, the gas was used for two EDM cell fillings with different pressures and polarizations. The first had higher pressure ($\sim 1$~bar) and lower polarization, and the second had lower pressure ($\sim 0.5$~bar) and higher polarization. After the EDM cell was filled, it was transported to the magnetically-shielded room in a battery-powered $400~\mathrm{\mu T}$ shielded solenoid. 
After the cell was placed in the measurement position, the BMSR-2 door was closed, and the magnetic field was allowed to stabilize for about five minutes. A time-dependent magnetic field along the $x$-axis with resonant frequency components and amplitudes tuned to effect a $\pi/2$ pulse for both species initiated each spin-precession run. Data were acquired from the $Z_1$-SQUID, which was located $(51\pm 1)$~mm above the center of the EDM cell. A silicon wafer was placed between the cell and dewar as indicated in Fig.~\ref{apparatus} to protect the SQUIDs from potential high voltage discharges.

The data-acquisition sample rate of 915.5245~Hz was derived from the 10~MHz output of an external clock~\cite{quartzclock}. % which has a relative stability of $<10^{-11}$ over 400~seconds integration time.%, which is much better than required for the comagnetometer data acquisition because clock instability is common mode for timescales longer than the 20 second block interval. 
The initial amplitudes of the precession signals were about 30~pT and 5~pT for $^{129}$Xe and $^3$He, respectively. The noise measured by the SQUID system in the BMSR-2 was $\sim 6~\mathrm{fT/\sqrt{Hz}}$ over the range  spanning the $\Xe$ and $\He$ frequencies. The free-precession decay time  $T_2^*$ was in the range of 3700--8000~s for $\Xe$ and 4000--8000~s for $\He$. Each run lasted about 15,000~s. 

As the $\Xe$ and $\He$ precessed, $\pm 6~$kV high voltage (HV) was applied to one electrode with the other electrode at ground potential. The average electric fields were $3.2~$kV/cm and $2.7~$kV/cm across cells PP1 and PP2, respectively. The relative uncertainty of the electric field, determined from modeling the cell and electrode geometry, was estimated to be 10\%. The voltage and field were chosen to be safely below the voltage observed to cause  breakdown across the cells at the lowest gas pressure used for this experiment. 

During each run, the HV polarity was positive ($+$), negative ($-$), or zero for equal intervals called segments, applied in a pattern that compensated drifts of comagnetometer frequency discussed below~\footnote{The 16-segment pattern $(+--+\ \ -++-\ \ -++-\ \ +--+)$ and its inverse compensated drifts accurately parametrized by a polynomial of up to \nth{3} order for equal length segments.}. For systematic studies, segments with zero HV were inserted at the beginning and end of each set of 16 segments within a run~\cite{Sachdeva2019}, and the rate of change of HV between segments (HV ramp) set to either 1 kV/s or 2 kV/s. Segments lengths of 400 or 800 seconds long were chosen based on the Allan deviation minimum from studies before taking EDM data and additional studies of frequency drifts during the experiment. The only data-analysis cuts were the shortening of eight out of a total of 539 segments due to HV or SQUID problems and five additional segments due to comagnetometer drift. 

The raw time-domain SQUID data were processed as follows: the DC offset and slow baseline drift were removed with an equiripple linear-phase finite-impulse-response high-pass filter with a stopband-edge frequency of 0.5~Hz and a passband frequency of 5~Hz. Filtered data were divided into non-overlapping blocks of length $\tau = 20~$seconds, chosen to be short enough that amplitude decay and frequency drift were negligible. Data for each block were fit over the interval $-\tau/2 \le t\le \tau/2$, with six free parameters, to the function
\begin{eqnarray}\label{fitmodel}
S(t) &= &a_{\mathrm{Xe}}\sin{\omega^\prime_{\mathrm{Xe}} t} + b_{\mathrm{Xe}}\cos{\omega^\prime_{\mathrm{Xe}} t}  \nonumber \\
& + &a_{\mathrm{He}}\sin{\omega^\prime_{\mathrm{He}} t} + b_{\mathrm{He}}\cos{\omega^\prime_{\mathrm{He}} t}.
\end{eqnarray}
The phase of each species at the middle of block $m$ was  $\Phi^m_{\mathrm{Xe/He}}= \arctan{(b^m_{\mathrm{Xe/He}}/a^m_{\mathrm{Xe/He}})}$$+2\pi N_m$, where $N_m$ is the integer number of cycles accumulated prior to the block. The uncertainty of $\Phi^m_{\mathrm{Xe/He}}$ was estimated from the covariance matrix of the fit scaled by the mean-square-error of the residuals. An alternative approach, which did not use the high-pass filter but added an offset and linear drift term to the fit function, produced consistent results. 

Magnetic field drifts were compensated by the comagnetometer corrected phases $\Phi^m_\mathrm{co} = \Phi^m_{\mathrm{Xe}} - R\Phi^m_{\mathrm{He}}$, where $R=1/2.7540816$ is the nominal ratio of the shielded gyromagnetic ratios of $\Xe$ and $\He$~\cite{Fan2016}. For each HV segment, the comagnetometer frequency $\omega_\mathrm{co}$ and uncertainty %$\sigma_{\omega_\mathrm{co}}$ 
were determined from the slope of a linear least-squares fit to $\Phi^m_{\mathrm{co}}$ as a function of time. The frequency uncertainties in all cases were consistent with the minimum expected uncertainties for constant-amplitude signals based on the signals, noise, and segment duration~\cite{Chupp1994,Chibane1995,Sachdeva2019}. Segment frequencies were blinded by adding or subtracting, depending on the sign of $\vec E\cdot\hat B$, an unknown frequency $\omega_\mathrm{blind}$ derived from a previously computer-generated pseudorandom number such that $|\omega_\mathrm{blind}|/(2\pi)\le$ 50~nHz. $\omega_\mathrm{blind}$ was saved separately from the data in a binary format. After all cuts and systematic corrections were determined, $\omega_\mathrm{blind}$ was set to zero to produce the set of HV segment frequencies for the final EDM analysis.

The EDM frequency was determined from an average of four consecutive segment frequencies with HV $(+--+)$ or $(-++-)$.
This compensated for linear drifts of the comagnetometer frequencies, typically  a few $\mu$Hz over the course of a run, which are predominantly due to effects of residual longitudinal magnetization that have been recently extensively studied~\cite{Terrano2018,Limes2018}. Higher order drifts produced a systematic error and were corrected as discussed below. 
%Repeated: Sequences of eight and 16 consecutive segment frequencies were used to study systematic effects of uncompensated drifts, as discussed below. 

Systematic effects include the uncertainties of experimental parameters as well as false-EDM signals that may arise from the nonideal response of the comagnetometer. The comagnetometer frequency $\omega_\mathrm{co}$ can be described by the following four dominant terms plus the EDM contribution $\omega_d\equiv\omega_{d_\mathrm{Xe}}-R\omega_{d_\mathrm{He}}$:
\begin{eqnarray}\label{comagmodel}
\omega_\mathrm{co}&\approx\ &\omega_{d}-\gamma^\prime_\mathrm{He}\Delta R B + \left (1-R \right)\vec\Omega\cdot\hat B \nonumber\\
&+& \gamma^\prime_\mathrm{Xe}\left (\Delta B^\mathrm{dif}_\mathrm{Xe}-\Delta B^\mathrm{dif}_\mathrm{He}\right )
+  \left (\omega^{sd}_\mathrm{Xe}-R\omega^{sd}_\mathrm{He} \right ).
\end{eqnarray}
Here, $\gamma^\prime_\mathrm{He/Xe}$ are the shielded gyromagnetic ratios; $\Delta R$ is a correction to $R$ that changes from run to run due mostly to pressure-dependence of the chemical shifts; $\vec B$ is the average magnetic field within the cell with contributions from the applied magnetic field $\vec B_0$, the ambient magnetic field of the room, and any nearby magnetized materials; $\vec\Omega$ is the angular frequency of the Earth's rotation; and $\Delta B^\mathrm{dif}_\mathrm{Xe/He}$ represents the difference of the volume average of the magnetic field and the field averaged by the atoms of each species as they move throughout the cell with different diffusion constants. In the presence of second- and higher-order gradients, this average is different for the two species~\cite{Sheng2014}. 

The $\nth{2}$ through $\nth{4}$ terms in Eq.~\ref{comagmodel} indicate the residual sensitivity of $\omega_\mathrm{co}$ to the magnitude, direction, and gradients of the magnetic field, and any correlation of these with the HV may cause a false-EDM signal. Such correlations are expected from possible leakage currents that flow across the cell, magnetization induced by charging currents that flow when the HV is changed, and motion of the measurement cell due to electrostatic forces that change with the HV. Our approach to estimating false-EDM signals is based on auxiliary measurements that measure the correlations with amplified leakage and charging currents, gradients, and cell motion, which are scaled to the HV correlations of these parameters monitored during the experiment. The last term in Eq.~\ref{comagmodel} reflects time-dependent, species-dependent shifts that dominate the comagnetometer drift~\cite{Terrano2018,Limes2018}. Eq.~\ref{comagmodel} does not include $\vec E \times \vec v$ effects, which are negligible. 

\begin{table}[t] %add [H] placement to break table across pages
\begin{ruledtabular}
\centering
\begin{tabular}{l c}
 \textbf{Source}  & \textbf{Sys. Error ($e\,\mathrm{cm}$)} \\ \hline
Leakage current &  $1.2 \times 10^{-28}$ \\ 
Charging currents & $1.7\times 10^{-29}$\\ 
$\vec E$-correlated cell motion (rotation)&$4.2\times10^{-29}$ \\ 
$\vec E$-correlated cell motion (translation)&$2.6\times10^{-28}$ \\ 
Comagnetometer drift & $6.6\times10^{-28}$ \\
$|\vec E|^2$ effects &  $1.2\times10^{-29}$\\ 
$|\vec E|$ uncertainty & $2.6\times10^{-29}$  \\
Geometric phase & $\le 2\times10^{-31}$ \\ 
 \hline
Total & $7.2\times10^{-28}$ 
\end{tabular}
\end{ruledtabular}
%\vskip -0.1 truein                     
\caption{\label{systematicstable} Summary of false EDM and other systematic effects discussed in the text.}
\vskip -0.1 truein
\end{table}
Systematic effects, including false EDM contributions and their uncertainties, are listed in Table \ref{systematicstable}. The leakage current was measured along the return path from the EDM cell electrode to the HV power supply with a maximum observed current of 100~pA. The comagnetometer response to such a current, simulated by a single turn of wire wrapped around the cell, was $\frac{1}{2\pi}\frac{\partial \omega_\mathrm{co}}{\partial I_{leak}} =(1.32 \pm 0.93)~\mathrm{\mu  Hz}/\mathrm{\mu A}$. Combining the two measurements, we determined a false EDM $(0.9\pm0.6)\times 10^{-28}~e\,\mathrm{cm}$ and, since the leakage current followed an unknown path that could increase or decrease $B$, we estimated the upper limit on the magnitude of a false EDM of $1.2\times 10^{-28}$~$e\,\mathrm{cm}$. During each HV ramp, the charging current might have induced magnetization of materials in or near the cell, correlated with the change of HV. The comagnetometer response to large charging currents was measured in studies with ramp currents of $\pm10~\mu$A and $\pm 20~\mu$A using the EDM-measurement HV pattern~\cite{Note1}. The measured comagnetometer frequency shift of $(-0.3 \pm 1.2)~\mathrm{nHz}/\mathrm{\mu A}$ combined with the maximum charging current of 19~nA measured during EDM data runs resulted in the upper limit on a false EDM of $1.7\times10^{-29}~e\,\mathrm{cm}$. 

The electric force between the cell electrodes and the grounded safety electrode might have caused the cell to move when the electric field was changed, affecting the magnetic fields and gradients across the cell. The effect of cell rotation on the comagnetometer frequency was measured by rotating the cell $\pm~5^\circ$ around the $z$-axis. HV-correlated cell rotation was investigated by measuring the motion of a laser beam spot reflected from the cell electrode with a lever arm of $1.5~$m and estimated to be less than 33 $\mu$rad. From these measurements, we determined the upper limit on the false EDM of $4.2\times10^{-29}~e\,$cm. HV-correlated translation of the cell in a non-uniform magnetic field might produce a false EDM because of the change of $B$ in the cell ($\nth{2}$ term in Eq.~\ref{comagmodel}) or through a change of the higher order gradients ($\nth{4}$ term in Eq.~\ref{comagmodel}). The $B$ dependence of $\omega_\mathrm{co}$ was estimated from the change in chemical shift for the different cell pressures. The HV-correlated amplitude of the spin-precession signals was used to estimate a limit of 30~$\mu$m on cell translation with respect to the SQUIDs. Combined with the linear magnetic field gradient based on $T_2^*$ for the two species~\cite{McGregor1990}, we determined the upper limit on the false EDM due to the translation in a linear gradient of $1.9\times 10^{-30}$~$e\,$cm. Another study combining an auxiliary measurement of the dependence of $\omega_\mathrm{co}$ on the current in a loop mounted on a cell electrode and the averaged field in the cell measured by the shift in the $^3$He frequency effectively isolated the $\nth{4}$ term in Eq.~\ref{comagmodel}. This provided an upper limit on any HV correlated effect, including cell motion, due to a source of magnetic field gradient outside the cell, provided the size of the source was smaller than its distance from the cell. The correlation $\frac{1}{2\pi}\frac{\partial \omega_\mathrm{co}}{\partial \omega_\mathrm{He}}=(-1.55\pm 0.28)\times 10^{-3}$ was then combined with the measured HV-correlated shift ${\delta\omega_\mathrm{He}}/{(2\pi)}=-(181.4\pm 124.4)$~nHz to set an upper limit of the false EDM due to cell translation of $2.6\times 10^{-28}$~$e\,$cm. Our future $\Xe$ EDM measurements will include a direct measurement of the dependence of $\omega_\mathrm{co}$ on cell translation and interferometric cell motion measurement.

Uncompensated drift of $\omega_\mathrm{co}$ would appear as a false EDM due to the frequency shift between segments with opposite $\vec E\cdot \hat B$. The comagnetometer frequency drifts for all runs were monotonic, and the time dependence could be accurately parametrized by polynomials of \nth{1} through \nth{4} order depending on the size of the drift and the signal-to-noise ratio, which varied from run to run. Offsets and linear drifts were compensated by the four-segment HV reversal pattern, while drifts characterized by \nth{2} and \nth{3} order time dependence were removed by the eight and 16-segment HV patterns, respectively. Because the linear time dependence is dominant, we have chosen to extract the EDM for four-segment measurements ($+--+$ or $-++-$) and to apply a correction for quadratic and higher order time dependence. The correction was calculated from the weighted polynomial coefficients of the fits to the comagnetometer frequency drift for each run. The polynomial order that accurately parametrized each run was determined by applying an $F$-test. We studied the dependence of the correction on the threshold $F_\mathrm{min}$ for $\int_{F_\mathrm{min}}^\infty P(F)dF$ and found corrections smaller than the uncertainty due to the fit parameters in all cases. For a threshold of 0.6, the correction was $(-0.08\pm 0.66)\times 10^{-27}$~$e\,\mathrm{cm}$, where the uncertainty is a statistical error based on the polynomial fits to the segment frequencies for each run, but is compiled as a systematic error in Table~\ref{systematicstable} to emphasize that it may give rise to a false EDM. As checks, the four-segment EDM result was compared to eight and 16-segment EDM measurements and found to be consistent.  A detailed description of comagnetometer drift and correction is presented in~\cite{Sachdeva2019,HeXeSupplement}.

$|\vec E|^2$ effects included any shift that depended on the magnitude of the applied electric field, for example, chemical shifts or HV-induced noise detected by the SQUID. Segments with $E=0$ and the different $E$ for the two cells enabled studies of the correlation of comagnetometer frequency with $\abs{E}$ and $|\vec E|^2$, providing an upper limit on the false EDM of $1.2\times10^{-29}~e\,\mathrm{cm}$. The modeling of the average electric field in the cell in the presence of the protection electrode contributed an uncertainty of $0.1d_A$. The combination of $\vec E\times \vec v$ effects coupled with magnetic field gradients could produce a false EDM, often referred to as a geometric phase. In gases at the densities used for these experiments, the time between collisions is small compared to the spin-precession period, which mitigates the coherent build up of a phase linear in the electric field. Using the formalism of Ref.~\cite{Pignol2015}, we found the false EDM due to geometric phase was $\leq 2\times 10^{-31}~e\,$cm. 

\begin{figure}[!t]
\centering
\includegraphics[width=\columnwidth]{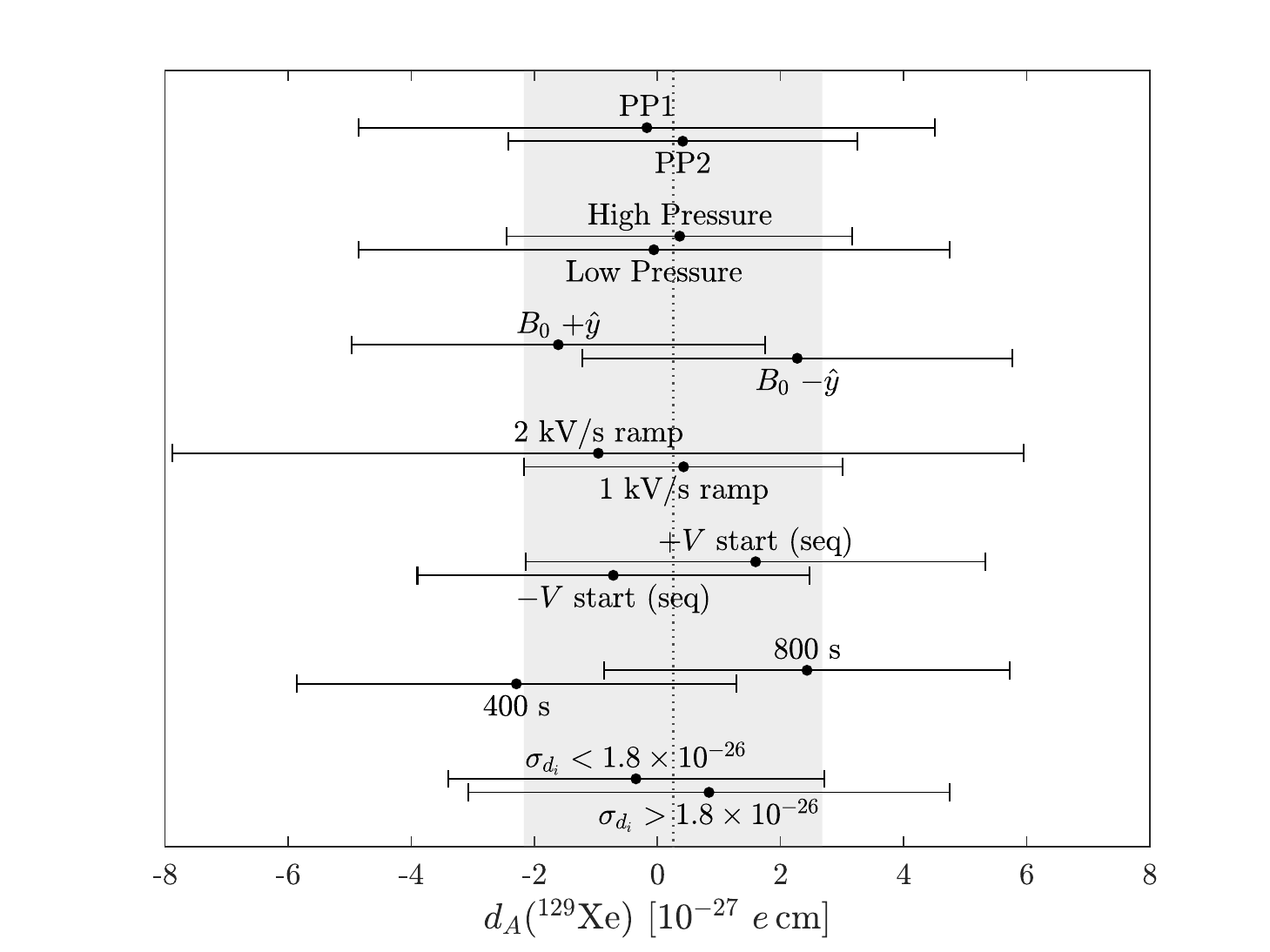}
\vskip -0.1 truein
\caption{\label{correlationsfig} Comparison of EDM measurements grouped by cell, cell pressure, $\hat B_0$ direction, HV ramp rate, HV start polarity, HV segment length, and an EDM uncertainty cut for $\sigma_{d_i}<1.8\times 10^{-26}~e\,\mathrm{cm}$ (14 measurements) or $\sigma_{d_i}>1.8\times 10^{-26}~e\,\mathrm{cm}$ (106 measurements). The shaded area shows the result given in Eq.~\ref{result}.}
\vskip -0.195 truein
\end{figure}

\begin{figure}[tb] %8.6cm
\centering
\vskip 0.06 truein
\includegraphics[width=\columnwidth]{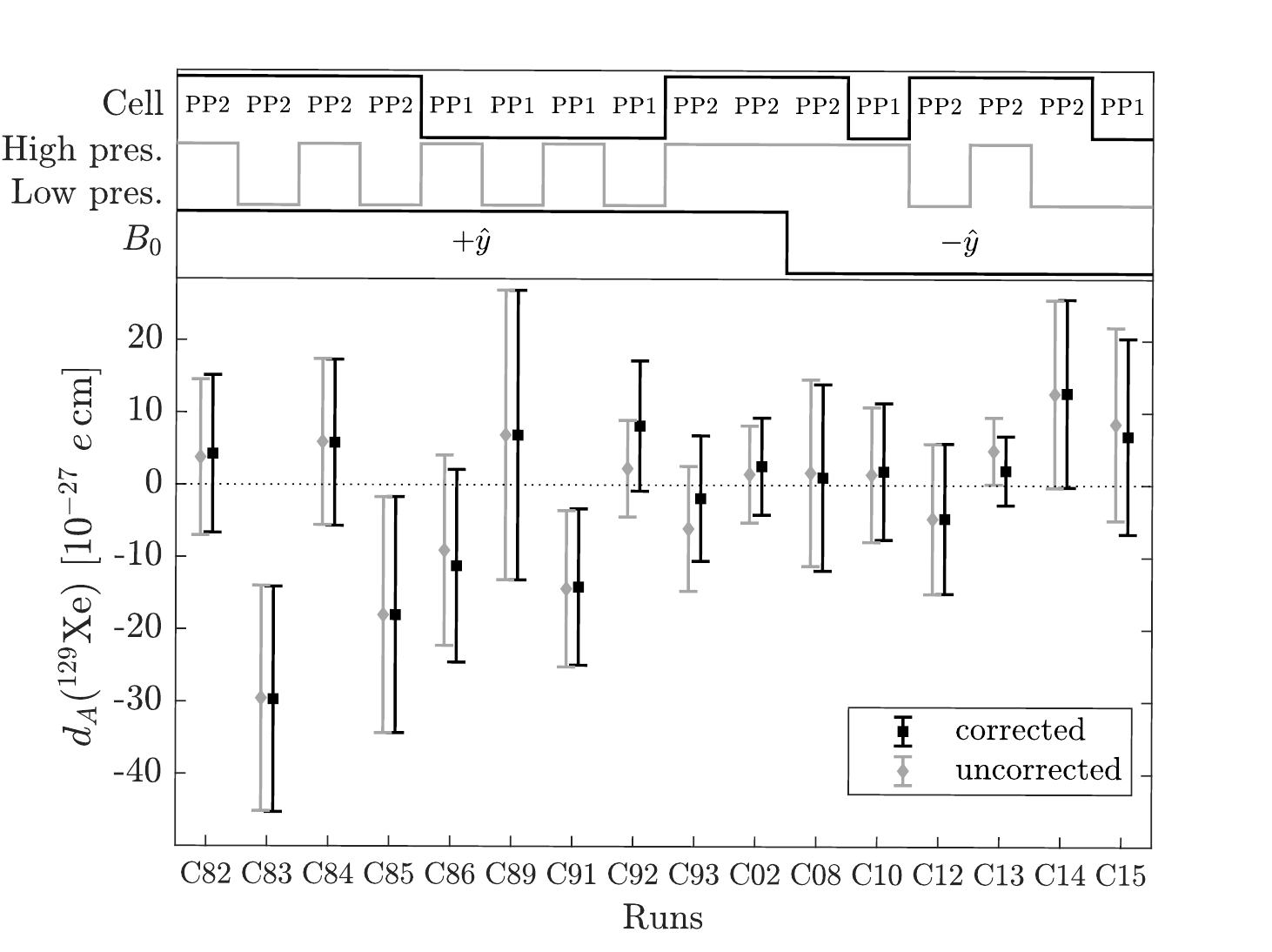}
\vskip -0.07 truein
\caption{\label{dataquality}  Weighted average of four-segment EDM measurements for each of the 16 runs indicating the cell used, cell pressure (high or low), and the magnetic field direction. For each run the uncorrected (left/gray) and drift-corrected (right/black) EDM are shown.}
%\vskip -0.15 truein
\end{figure}

A total of 120 EDM measurements were acquired in 16 separate runs under a variety of different conditions including measurement cell (PP1 or PP2), gas pressure, $\vec B_0$ direction, HV ramp rate, HV polarity at the start of the EDM measurement, HV segment length, and a cut on the EDM uncertainty $\sigma_{d_i}$. Fig.~\ref{correlationsfig} shows a comparison of sorting all EDM measurements into two groups based on these variables, and Fig.~\ref{dataquality} shows the EDM measurements combined into runs that had different cells, cell pressures, and orientations of $\vec B_0$. We also investigated correlations between the extracted EDM and other parameters including $T_2^*$ and an alternative combination of four segment frequencies weighted by $\pm(+1\, -3\, +3\, -1)$, which is even under reversal of HV and insensitive to linear and quadratic drifts. The absence of any correlations provided checks on additional effects that may have been related to differences of the HV ramp between the $\nth{1}$--$\nth{2}$ and the $\nth{3}$--$\nth{4}$ HV reversals. %A Monte Carlo study using the time-domain noise data from all runs with added simulated signals for $^{129}$Xe and $^3$He at slightly different frequencies confirmed that an EDM of $3\times 10^{-27}~e\,\mathrm{cm}$ could be resolved with our EDM uncertainty.

The consistency of EDM measurements over the variety of conditions and cuts illustrated in Fig.~\ref{correlationsfig} justified taking the weighted average of the EDM measurements, providing the comagnetometer-drift corrected result
\begin{equation}
d_A(\Xe) = (0.26 \pm 2.33\ (\mathrm{stat}))\times 10^{-27}~e\,\mathrm{cm}.
\label{result}
\end{equation}
The statistical error is the square root of the inverse of the sum of the weights of the uncorrected measurements, and $\chi^2=106.1$ for 119 D.F. Combined with the systematic error from Table~\ref{systematicstable}, we find $|d_A(\Xe)| \le 4.81\times 10^{-27}~e\,\mathrm{cm}$ (95\% CL). This is a factor of 1.4 improvement in sensitivity over the previous best result with one week of data compared to six months for the measurement of Ref.~\cite{Rosenberry2001}. 
% Distribution of weighted EDM values $d/\sigma_d$? $E^2$ plot of extracted frequencies? (Rosenberry did this).
%\begin{figure}[tb] %8.6cm
%\centering
%\includegraphics[width=\columnwidth]{histogram_bw}
%\vskip -0.14 truein
%\caption{\label{histogram} Histogram of the normalized residuals for all 120 EDM measurements, where $r_i=(d_i-d_A(\Xe))/\sigma_{d_i}$ for the weighted average of $d_A(\Xe)$ (Eq.~\ref{result}), $d_i$ is a single four-segment EDM measurement, and $\sigma_{d_i}$ is its uncertainty; the solid curve is a Gaussian with unit variance and area 120.}
%\vskip -0.14 truein
%\end{figure}
Significant improvements to the polarization, SQUID dewar noise, measurement time, and increased electric field should improve the $\Xe$ EDM sensitivity by an order of magnitude or more. The largest systematic error, due to comagnetometer drift,  can be reduced with more precise $\pi/2$ pulses and optimized EDM cell shape. Higher resolution leakage current measurement and improved cell motion measurements are also essential. While any further increase in sensitivity and upper limit will impact the global interpretation of EDM results \cite{Chupp2015}, an order of magnitude improvement would represent a significant advancement in our sensitivity to BSM physics. This work also presents significant advances in comagnetometer analysis \cite{HeXeSupplement} and may have impact on other EDM and BSM searches including planned neutron EDM experiments with comagnetometers.
\begin{acknowledgments}
We wish to thank Patrick Pistel and Roy Wentz for excellence and innovation in glass blowing and cell construction. This work was supported in part  by NSF grant PHY-1506021, DOE grant DE-FG0204ER41331, Michigan State University, by  Deutsche Forschungsgemeinshaft grants TR408/12 and FA1456/1-1 and The Cluster of Excellence "Origin and Structure of the Universe." WT acknowledges the support of a Humbolt Stiftung Fellowship.
\end{acknowledgments}

%\bibliography{references}
%\input{HeXe2017arXiv.bbl}


\end{document}